\documentclass[article]{aa}
\usepackage{natbib}
\usepackage{graphicx}
\usepackage{color}
\usepackage{sidecap}
\begin{document}{}

  \title{Simulated Magnetic Flows in the Solar Photosphere}
   \titlerunning{Simulated Magnetic Flows}
   \author{S.~Danilovic \inst{1} \and R.~H.~Cameron\inst{1} \and S.~K.~Solanki \inst{1,2}}

   \institute{Max-Planck-Institut f\"ur Sonnensystemforschung, Justus-von-Liebig-Weg 3
37077 G\"ottingen, Germany \and 
   School of Space Research, Kyung Hee University, Yougin, Gyeonggi 446-701, Korea}

   \date{\today}

\abstract
 {Recent Sunrise/IMaX observations have revealed the existence of supersonic magnetic flows.}
  {Our aim is to determine the origin of such flows by using realistic MHD simulations.}
  {We simulate cancellation and emergence of magnetic flux through the solar photosphere. Our first numerical experiment
starts with magnetic field of both polarities. To simulate emergence into a region with pre-existing field, we introduce a large-scale horizontally uniform 
sheet of horizontal field. We follow the subsequent evolution, creating synthetic polarimetric observations, including known instrumental 
effects of 
the Sunrise/IMaX and Hinode/SP instruments. We compare the simulated and observed spectropolarimetric signals.}
  {Strongly blue- and redshifted Stokes V signals are produced in locations where strong line-of-sight velocities coincide with the strong 
line-of-sight component of magnetic field. The size and strength of simulated events is smaller than observed and they are mostly associated 
with downflows, contrary to observations. In a few cases where they appear above a granule, single blue lobed Stokes V are produced due to 
strong gradients in magnetic field and velocity. No change of magnetic field sign is detected along the line of sight in these instances.
More high-speed magnetized flows occur in the case where emergence is simulated then in the case where no horizontal field was added.}
  {The simulations indicate that the observed events result from magnetic flux emergence, where reconnection may take place but does not seem to be necessary.}

\maketitle

\section{Introduction}

Advances in instrumentation often lead to the discovery of new phenomena in the solar atmosphere. Thus, \cite{Borrero:etal:2010} found a large number of strongly blue-shifted Stokes V profiles in observations recorded by the  Imaging Magnetograph eXperiment \citep[IMaX,][]{Valentin:etal:2011a} on board Sunrise balloon-borne solar observatory 
\citep{Barthol:etal:2011,Solanki:etal:2010,Berke:etal:2011}. By assuming hG vertical field and after taking the IMaX spectral resolution into account, \cite{Borrero:etal:2012} 
concluded that supersonic velocities are needed to produce such signatures at a wavelength shift of $\Delta\lambda=+227$~m\AA~ from the Fe~I~$525.02$~nm nominal line center. Since these features appear mostly above granules and thus are related to upflows, it is expected that the signal actually comes from the blueshifted neighbouring Fe~I~$525.06$~nm line.

Counterparts of these events were later found in Hinode/SP \citep{Lites:etal:2001,Kosugi:etal:2007} data. \cite{Valentin:etal:2011} assumed 
that Fe~I~$525.02$~nm and Fe~I~$630.25$~nm lines are similar and equally sensitive to all atmospheric parameters and identified signals 
that correspond to the same Doppler velocities. The difference  with respect to IMaX results is that, while IMaX found  $70$\% of cases to be associated with upflows, in Hinode data blue-shifted and red-shifted profiles were equally present. On the other hand, both datasets, as well as the detailed study by \cite{Carlos:etal:2013},
showed that the events coincide with the appearance of linear polarization in the regions where the sign of the line-of-sight field was mixed. Based on these characteristics, both groups of authors concluded that reconnection of the emerging with pre-existing field could be the cause 
of these magnetic flows. This is supported by \cite{Borrero:etal:2013} who found a reversal in the polarity of the magnetic field along the line-of-sight, accompanied by an enhancement in the temperature and supersonic line-of-sight velocities.
 
We explore this scenario by using numerical 3D radiation MHD simulations of an emerging magnetic flux sheet in the presence of magnetic flux concentrations. We compute spectral line profiles in the simulated snapshots to simulate the IMaX or Hinode observations and compare the results to the observed signatures.

\begin{figure}
\includegraphics[angle=-90,width=\linewidth,trim= 6.2cm 0cm 4cm 0cm,clip=true]{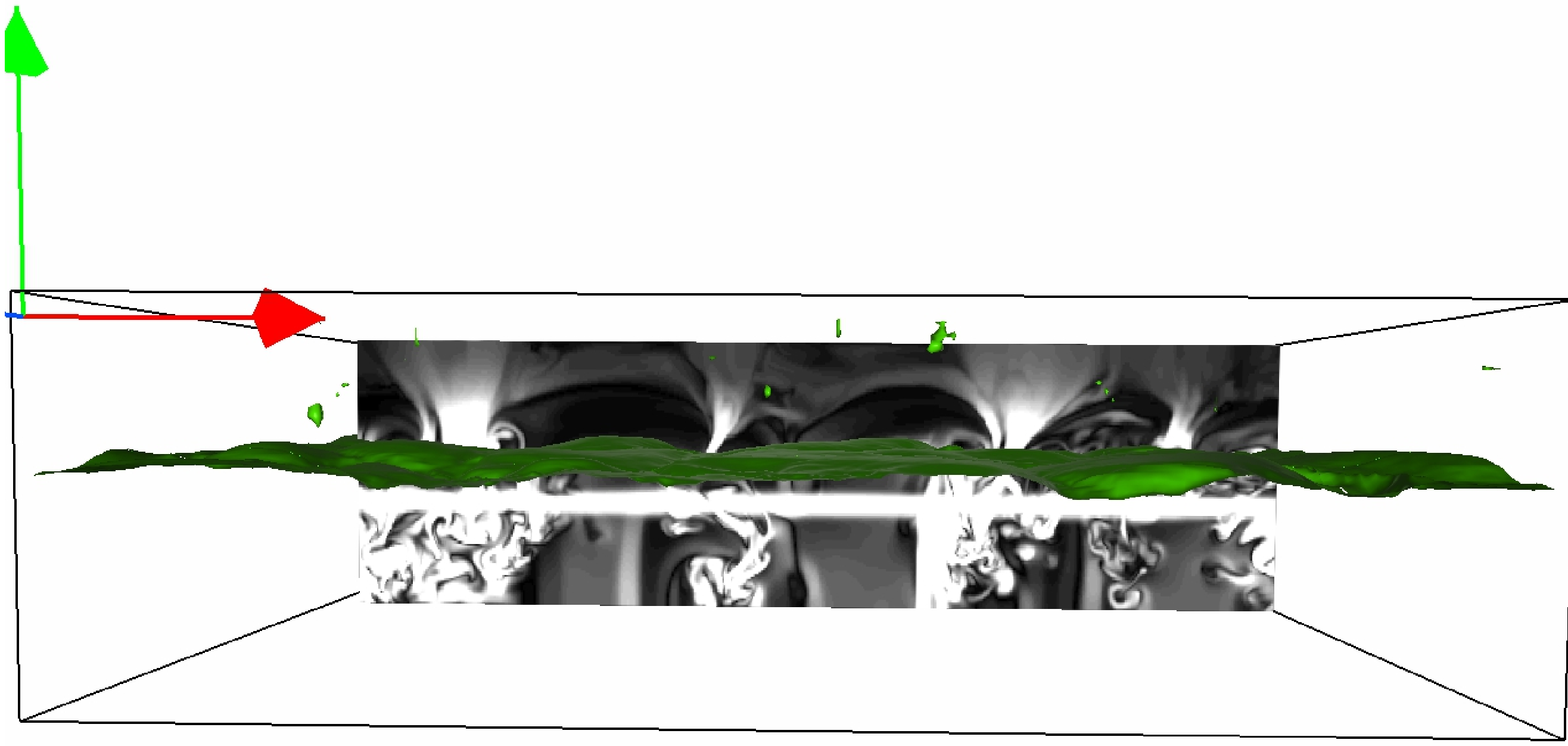}
\includegraphics[angle=-90,width=\linewidth,trim= 6.2cm 0cm 4cm 0cm,clip=true]{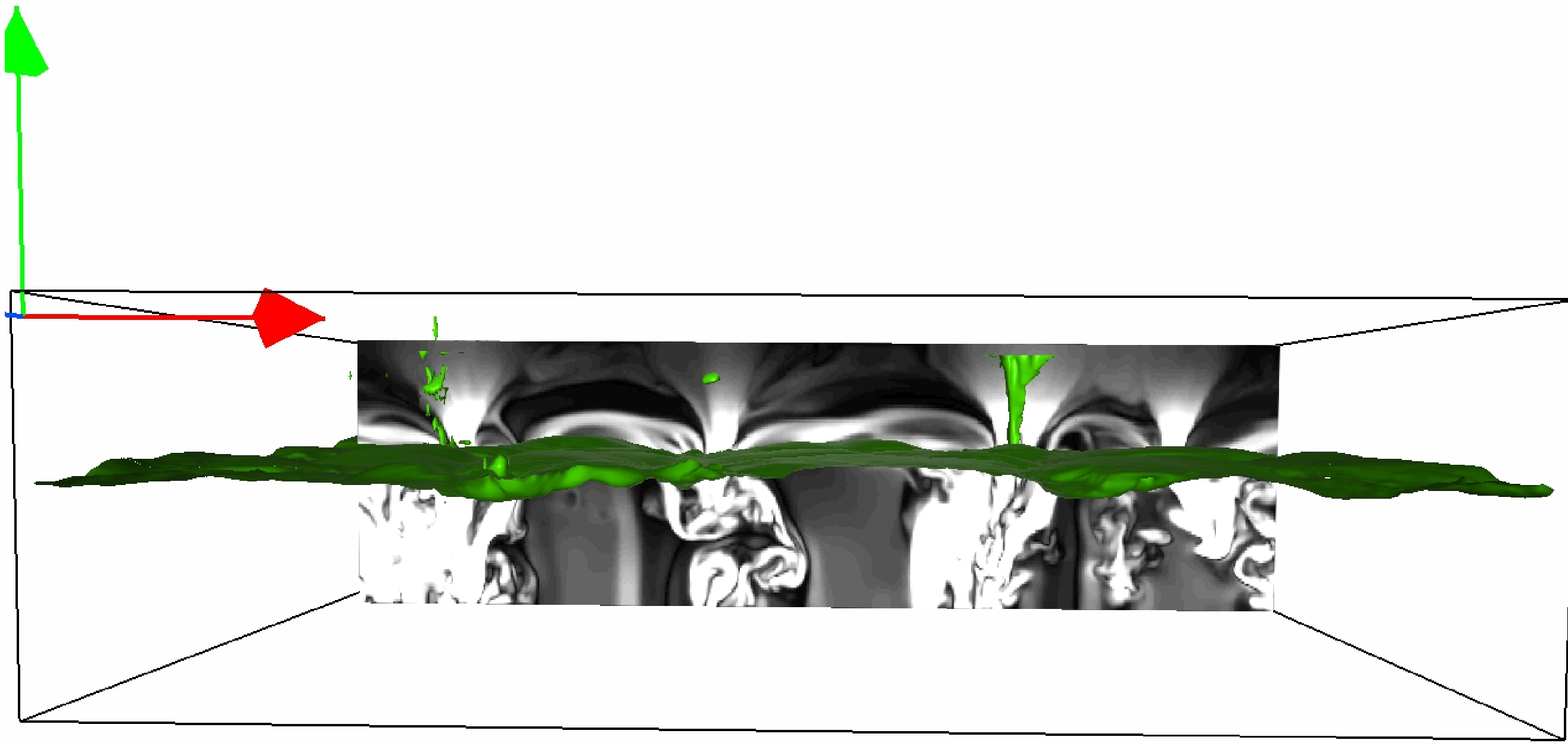}
\caption{Two snapshots of the 'emergence run' - side view of the simulation domain at t=0 s (upper) and t=188 s. Vertical plane shows magnetic field strength. The direction of the introduced horizontal field is shown with red arrows (+X direction). The green isosurface of a temperature of 6000 K marks the position of $\tau_{500}\approx 1$ level. Above this corrugated plane, green features are visible signalling hot gas in the upper part of the simulation domain. This is where reconnection takes place or shock fronts are formed. \label{fig:cuts}}
\end{figure}

\section{Computational runs and synthetic observations}

We consider two runs produced with the MURaM code \citep{Voegler:etal:2005}. The setup is essentially the same as in 
\cite{Cameron:etal:2011} with a checkerboard-like $2\times2$ pattern of initial magnetic polarity, vertical magnetic field boundary 
condition at the upper boundary which is transparent for upflows and non-grey radiative transfer included. 
The computational domain of $6\times6\times1.68$~Mm was chosen with some $700$~km above  $\tau_{500}=1$ 
and the resolution of $10$ and $14$~km in horizontal and vertical directions, respectively. 
Both ambipolar diffusion and the Hall current are included in the induction equation. 
These terms are expected to be important in the 
photosphere for dynamics at very small scales, especially in regions where the magnetic field gradients are large \citep{Mark:Robert:2012}. At the current resolution (10~km in the horizontal directions) the effects on the dynamics of quiet-Sun features 
are relatively weak \citep{Mark:Robert:2012}. A study of the full effects of the ambipolar and Hall effects is an interesting
topic for future studies at higher spatial resolution. 

Both runs start from the same snapshot of non-magnetic convection, to which a purely vertical magnetic 
field of positive and negative polarity was added. This initial field is arranged in a $2\times2$ checkerboard pattern
\citep[as in ][]{Cameron:etal:2011}, and has a height-independent strength of 200~G.
The first run (from now on called the 'reference run'), was allowed to evolve freely from this initial condition, 
simulating the decay of the magnetic field in a mixed polarity region. For the second run we introduced a horizontal flux sheet $300$~km below mean optical depth unity, in order to simulate the scenario explained in the introduction.
The field strength is set to vary across the cross section as a Gaussian with a FWHM of  $110$~km with a maximum value of $1$~kG. 
We call this run - the 'emergence run'. In what follows all times are given with respect to the time at which the flux
sheet was introduced in the second run.

Figure~\ref{fig:cuts} shows a vertical cut through the initial snapshot (with the imposed sheet of horizontal flux) and the snapshot 188~s later. In the upper panel, the sheet is unmodulated, in its initial position. The field strength of $1$~kG at the center of the sheet gradually decrease to $0$~G at the edges. 
The added field was directed in the +X axis, i.e. aligned with the planes shown in this figure.
At t$=188$~s most of the flux in this sheet has emerged. At this time, high-temperature structures are
visible in locations where the emerging flux is reconnecting with the pre-existing magnetic field (green features in upper part of the simulation domain). We stress that it is not our aim to follow the detailed emergence and evolution of individual loops or the sheet in general. Rather, we are interested in the interaction of  emerging flux with pre-exiting field in the solar photosphere. The number of such interactions is maximized with the initial magnetic configuration chosen here. Additionally, the time needed for the field to emerge is minimized.

To explore the observational signatures of these interactions, we synthesized the Sunrise/IMaX and Hinode/SP observables with the radiative transfer code SPINOR \citep{Frutiger:etal:2000}, based on the STOPRO routines \citep{Solanki:1987}. The Fe abundance used for the synthesis is taken from \citet{Thevenin:1989} and the values of the oscillator strengths from the VALD database \citep{Piskunov:etal:1995}. For simulated IMaX data we synthesized the  Fe~I~$525.02$~nm line, the Fe~I~$525.06$~nm line and the Co~I~$525.0$~nm line, which were then smeared with the IMaX spectral PSF. 
To simulate the effect of the phase diversity reconstruction technique with which the data were reconstructed that \cite{Borrero:etal:2010} used, we applied a low-pass filter that removes all spatial frequencies higher than $5.5$~arcsec$^{-1}$. We also took into account the stray-light which has been estimated empirically based on data recorded at the limb (Feller et al., in preparation). A noise of $3\cdot 10^{-3}$~I$_{c}$ was added. 

The Hinode/SP lines Fe~I~$630.15$ and $630.25$~nm were also computed. They were broadened with the Hinode/SP spectral PSF, as well as with a realistic spatial PSF \citep{Danilovic:etal:2008}. Finally, random noise of $8\cdot 10^{-4}$~I$_{c}$ was added.

\begin{figure*}
\includegraphics[angle=0,width=\linewidth,trim= 1cm 13cm 1cm 11cm,clip=true]{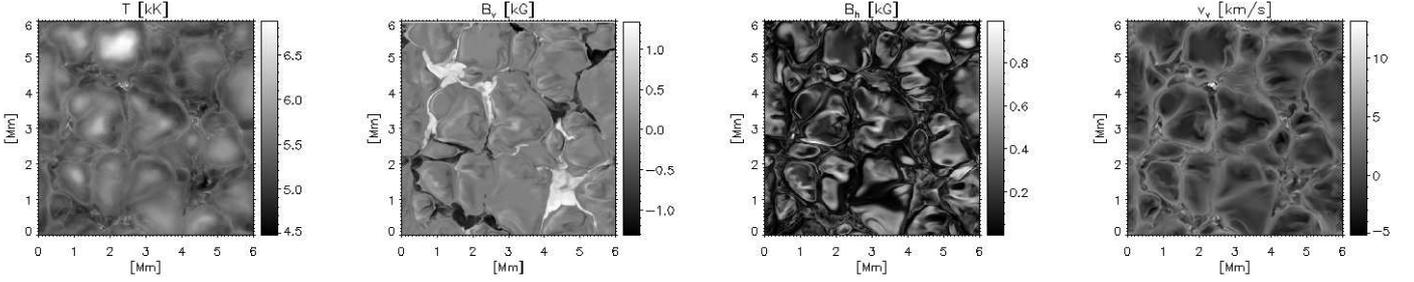}
\caption{'Emergence run' at t$=156$~s - horizontal cut through the simulation domain (for the evolution of these quantities see the on-line material)- \textit{left to right}: temperature, vertical and horizontal field, line-of-sight velocity (upflow is negative) at constant geometrical height that corresponds to $< \tau_{500} > \approx 1$. \label{fig:all_mhd}}
\end{figure*}

\begin{figure*}
\includegraphics[angle=0,width=\linewidth,trim= 1cm 13cm 1cm 11cm,clip=true]{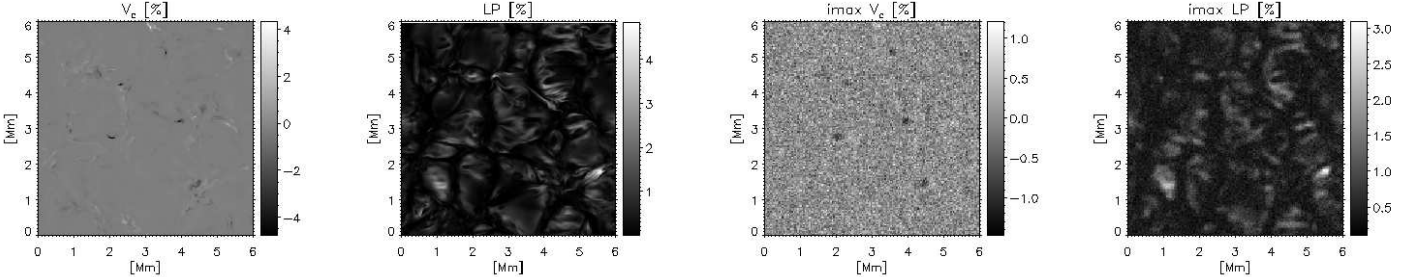}
\caption{Synthesized SUNRISE/IMaX observables from the 'emergence run' at t$=156$~s (for the evolution of these quantities see the on-line material)- \textit{left to right}: Stokes V $\Delta\lambda=+227$~m\AA~away from nominal Fe I $525.02$~nm line center and mean linear polarization at original and IMaX resolution. \label{fig:all_imax}}
\end{figure*}

\begin{figure*}
\includegraphics[angle=0,width=\linewidth,trim= 1cm 11.5cm 1cm 11cm,clip=true]{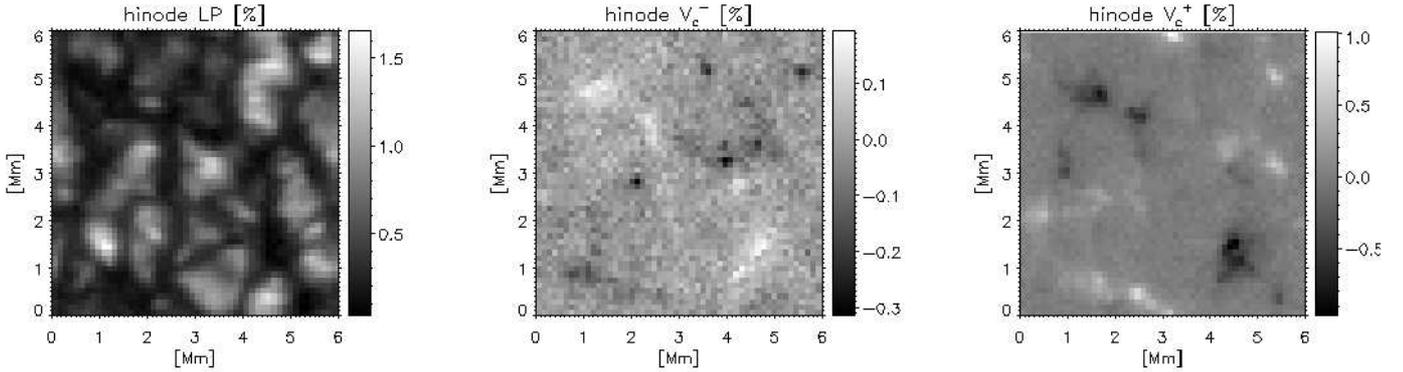}
\caption{Synthesized Hinode/SP observables from the 'emergence run' at t$=156$~s (for the evolution of these quantities see the on-line material)- \textit{left to right}: mean linear polarization and Stokes V at $\Delta\lambda=-227$~m\AA~and $+227$~m\AA~away from nominal Fe I $630.25$~nm line center. \label{fig:all_hinode}}
\end{figure*}

\begin{figure}
\includegraphics[angle=90,width=0.8\linewidth,trim= 3cm 0cm 0cm 0cm,clip=true]{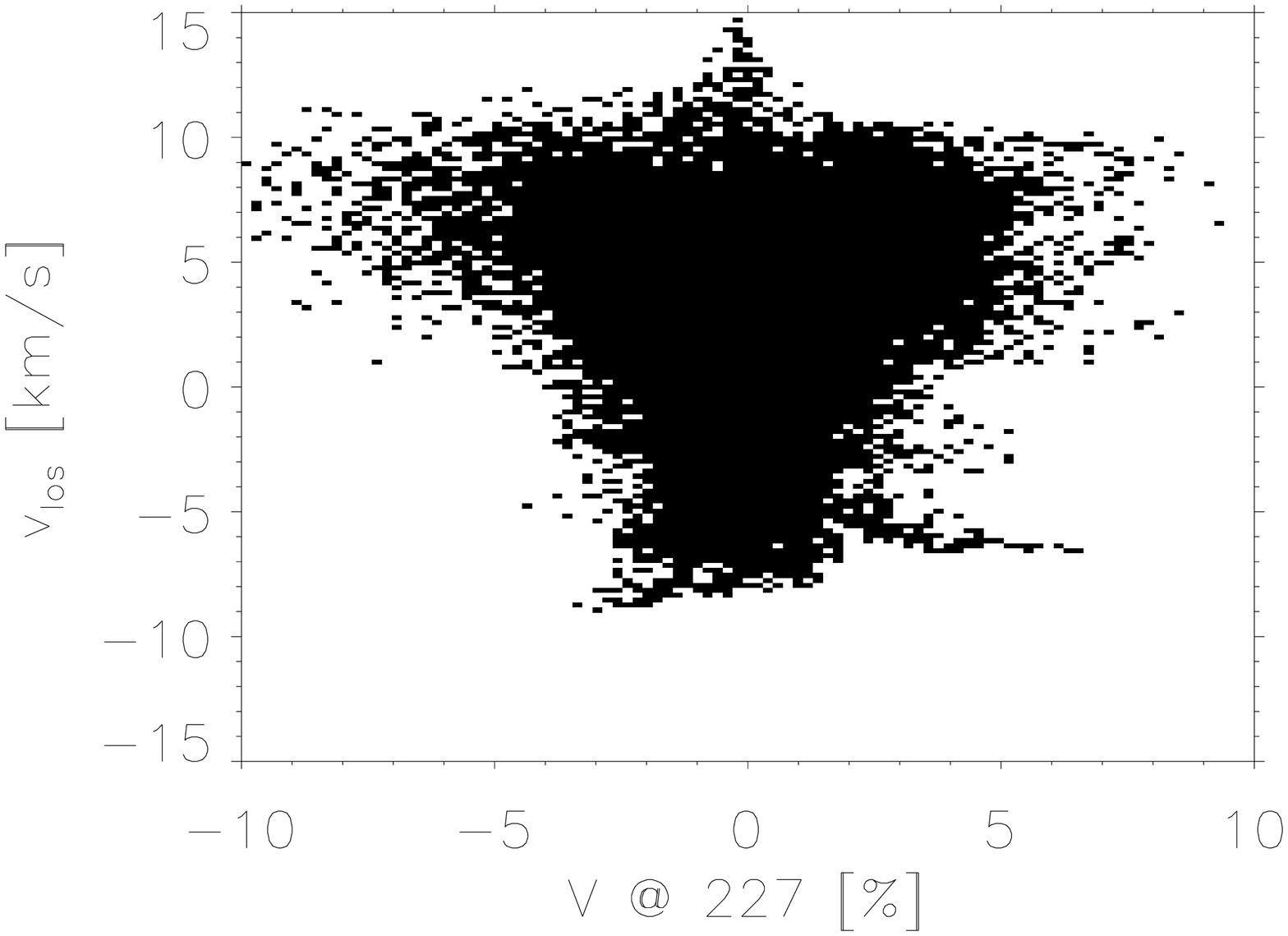}
\includegraphics[angle=90,width=0.8\linewidth,trim= 3cm 0cm 1cm 0cm,clip=true]{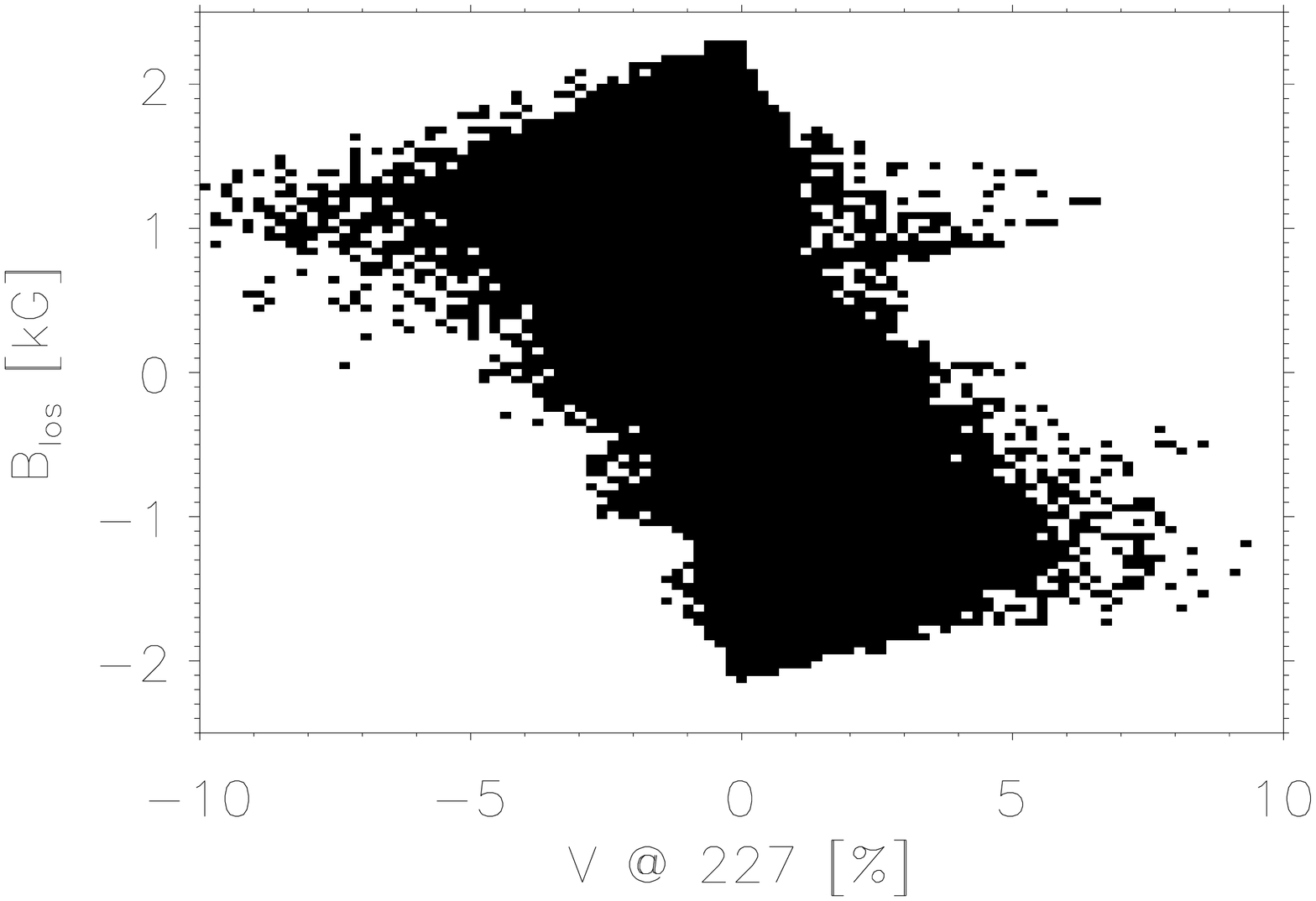}
\includegraphics[angle=90,width=0.8\linewidth,trim= 0cm 0cm 1cm 0cm,clip=true]{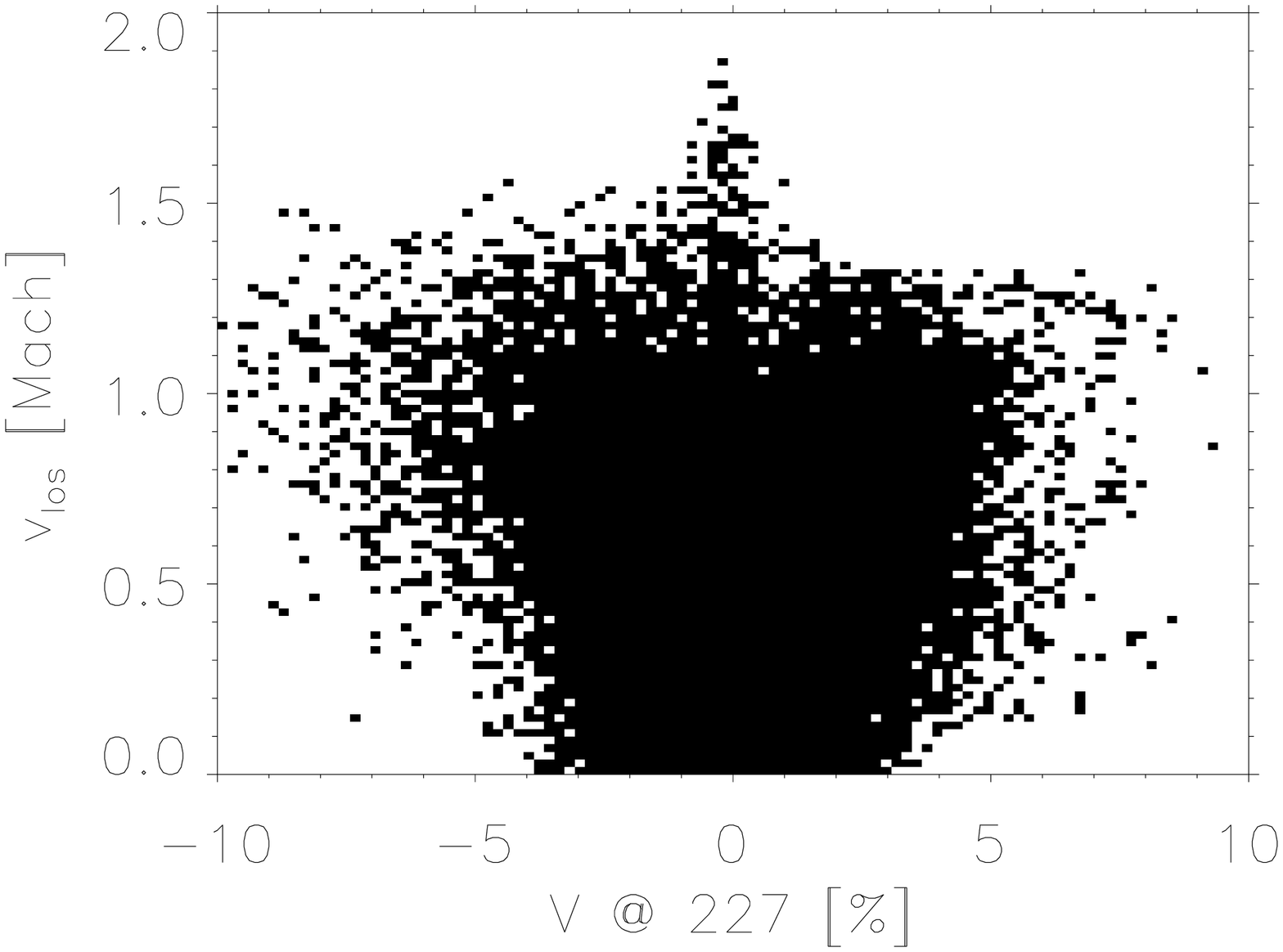}
\caption{Scatter plots of the synthesized IMaX Stokes V signal at $\Delta\lambda=227$~m\AA~ from Fe I 525.02 nm nominal line centre (i.e. 'continuum' Stokes V or V$^{-}_{c}$) versus line-of-sight velocity (top panel) and magnetic field (middle panel) at $\log\tau=-1$. Bottom panel shows line-of-sight velocity in units of the Mach number. \label{fig:2dhist}}
\end{figure}

\begin{figure}
\centering
\includegraphics[width=0.7\linewidth,angle=90,trim= 0cm 0cm 0cm 0cm,clip=true]{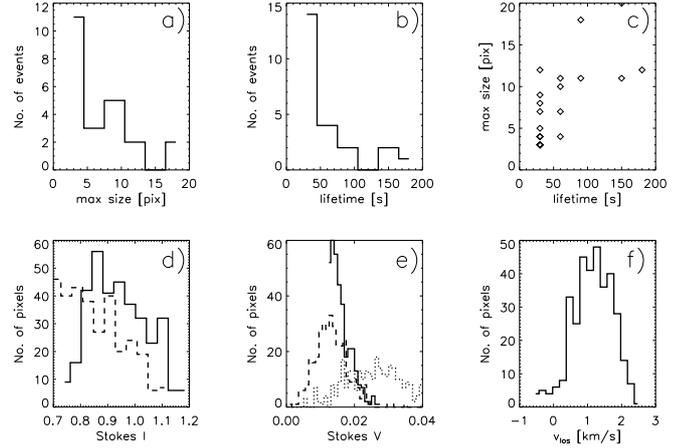}
\caption{Characteristics of selected simulated events: a) distribution of maximal area, b) lifetimes, c) relations between the area and the lifetime, d) corresponding Stokes I profiles (solid - continuum intensity distribution; dashed - line averaged intensity), e) corresponding Stokes V profiles (solid - Stokes V at $\Delta\lambda=+227$~m\AA ; dashed - mean, line-averaged Stokes V; dotted - maximum of Stokes V) and f) line of sight velocity retrieved from Stokes I. The figure  is comparable to Fig. 2 from \cite{Borrero:etal:2010} which shows the same parameters retrieved from observations. \label{fig:hist_obs}}
\end{figure}

\begin{figure*}
\centering

\includegraphics*[width=0.29\linewidth,angle=90,trim= 0cm 0cm 0cm 0cm,clip=true]{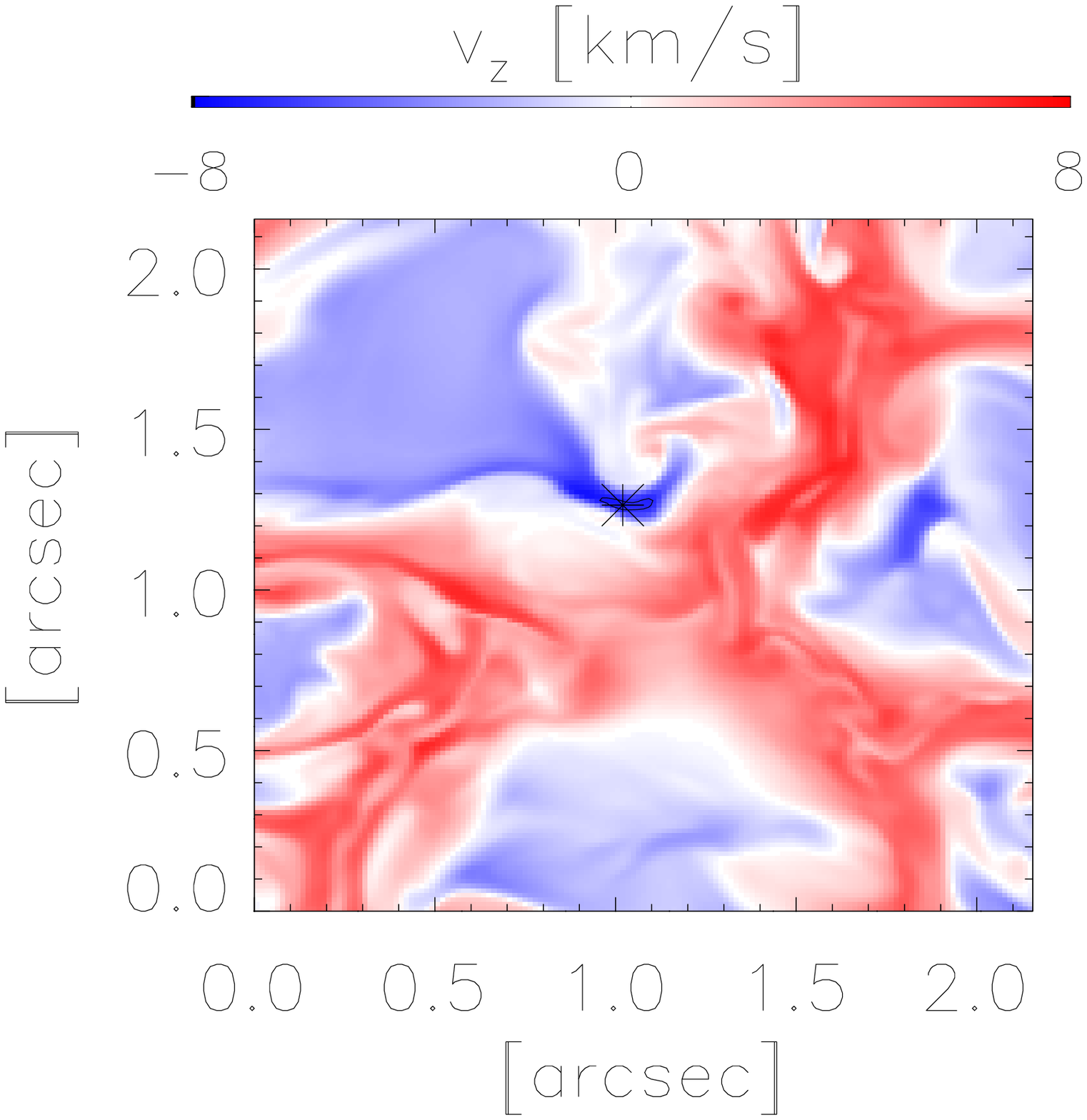}
\includegraphics*[width=0.29\linewidth,angle=90,trim= 0cm 0cm 0cm 0cm,clip=true]{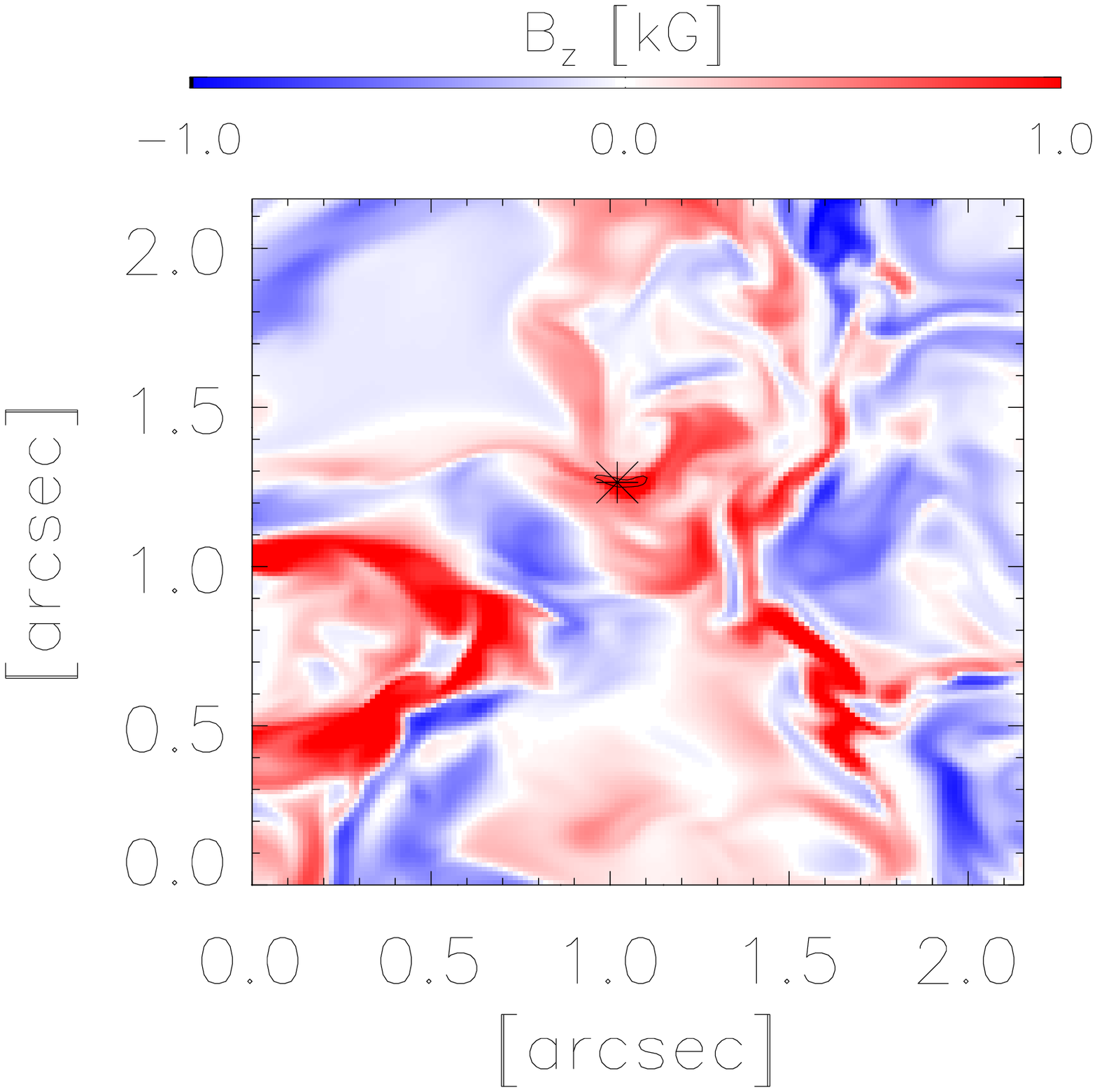}
\includegraphics*[width=0.29\linewidth,angle=90,trim= 0cm 0cm 0cm 0cm,clip=true]{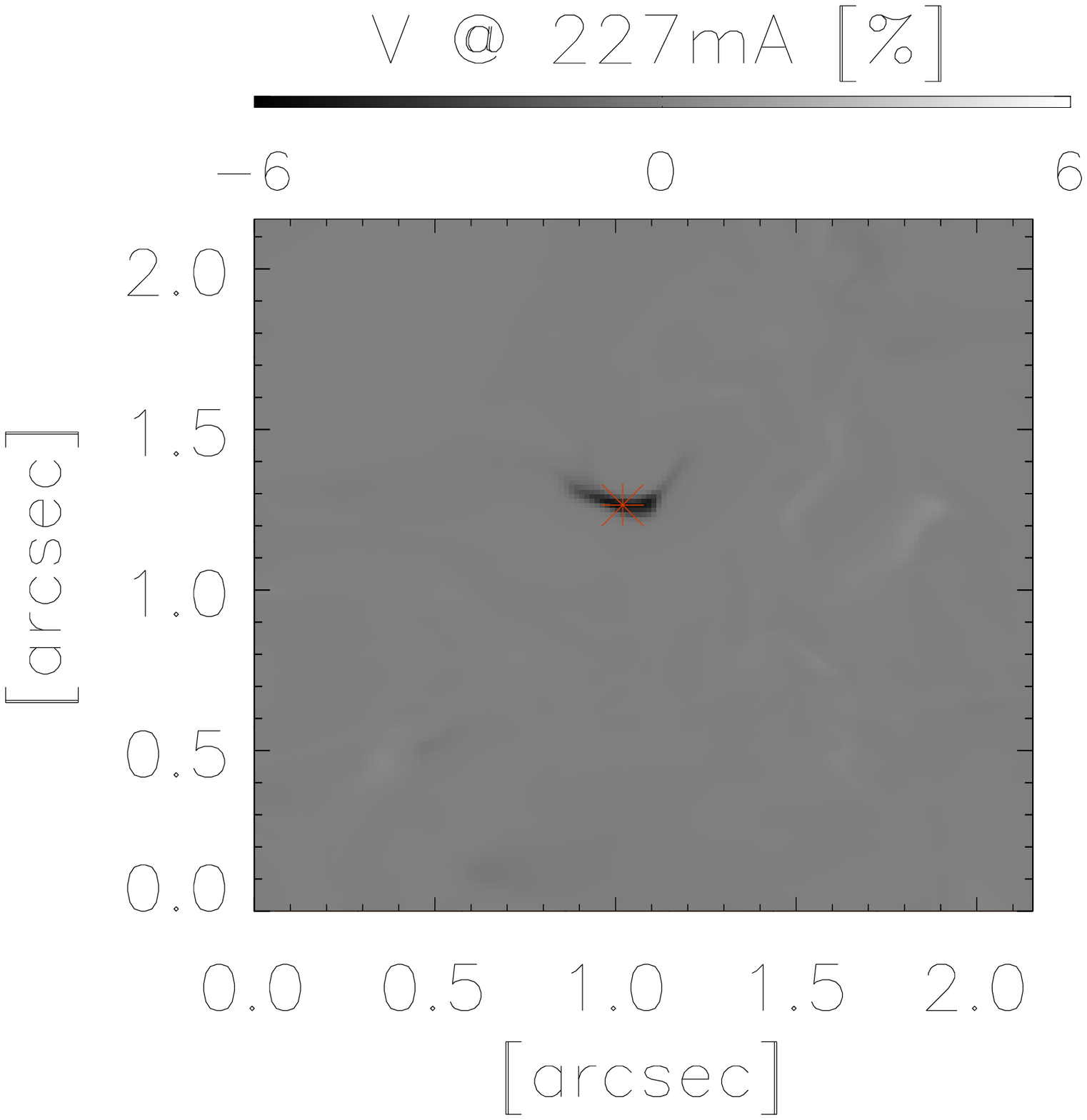}\\
\includegraphics*[width=0.175\linewidth,angle=90,trim= 1cm 1cm 1cm 4cm,clip=true]{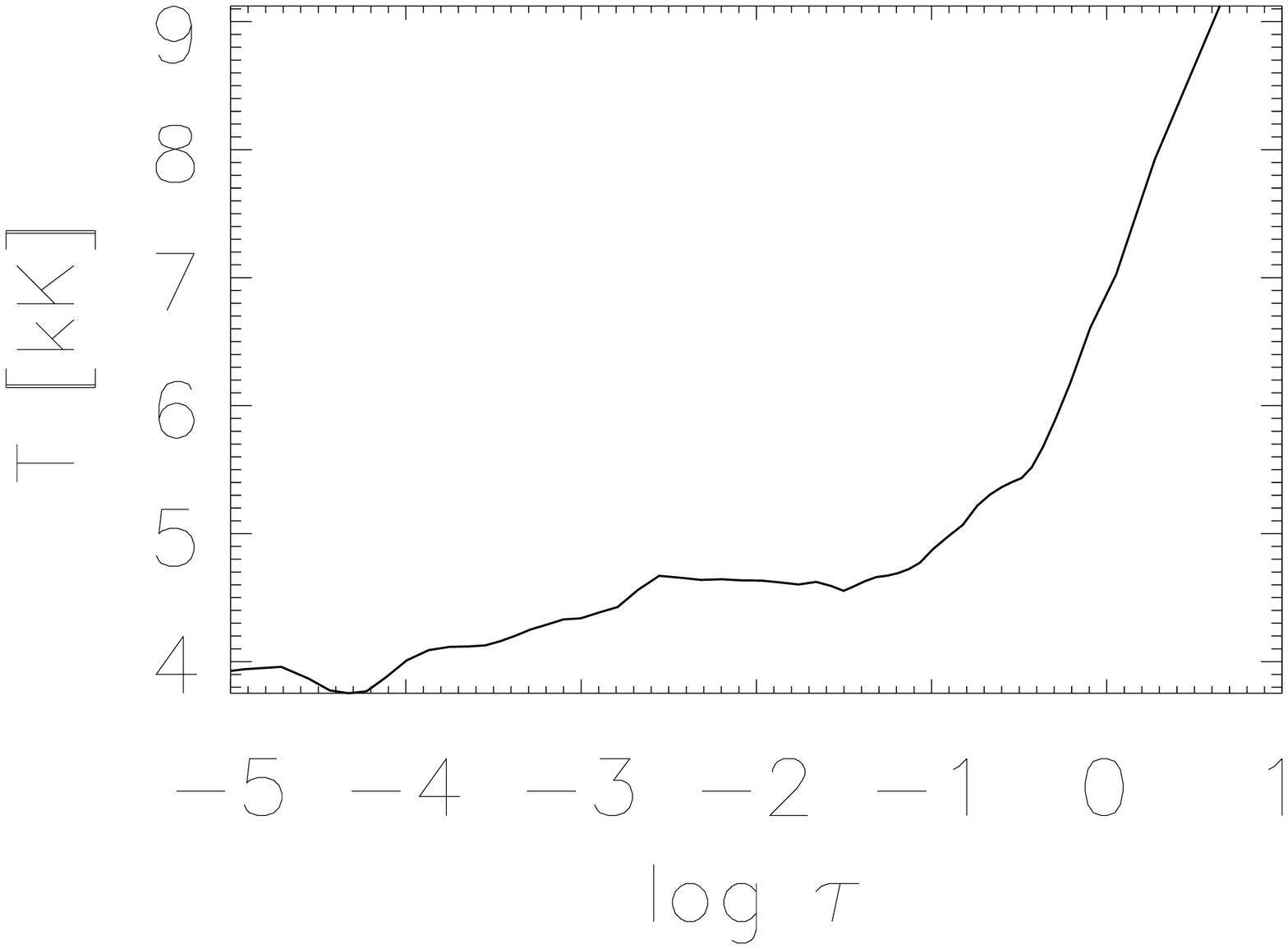}
\includegraphics*[width=0.175\linewidth,angle=90,trim= 1cm 1cm 1cm 3.3cm,clip=true]{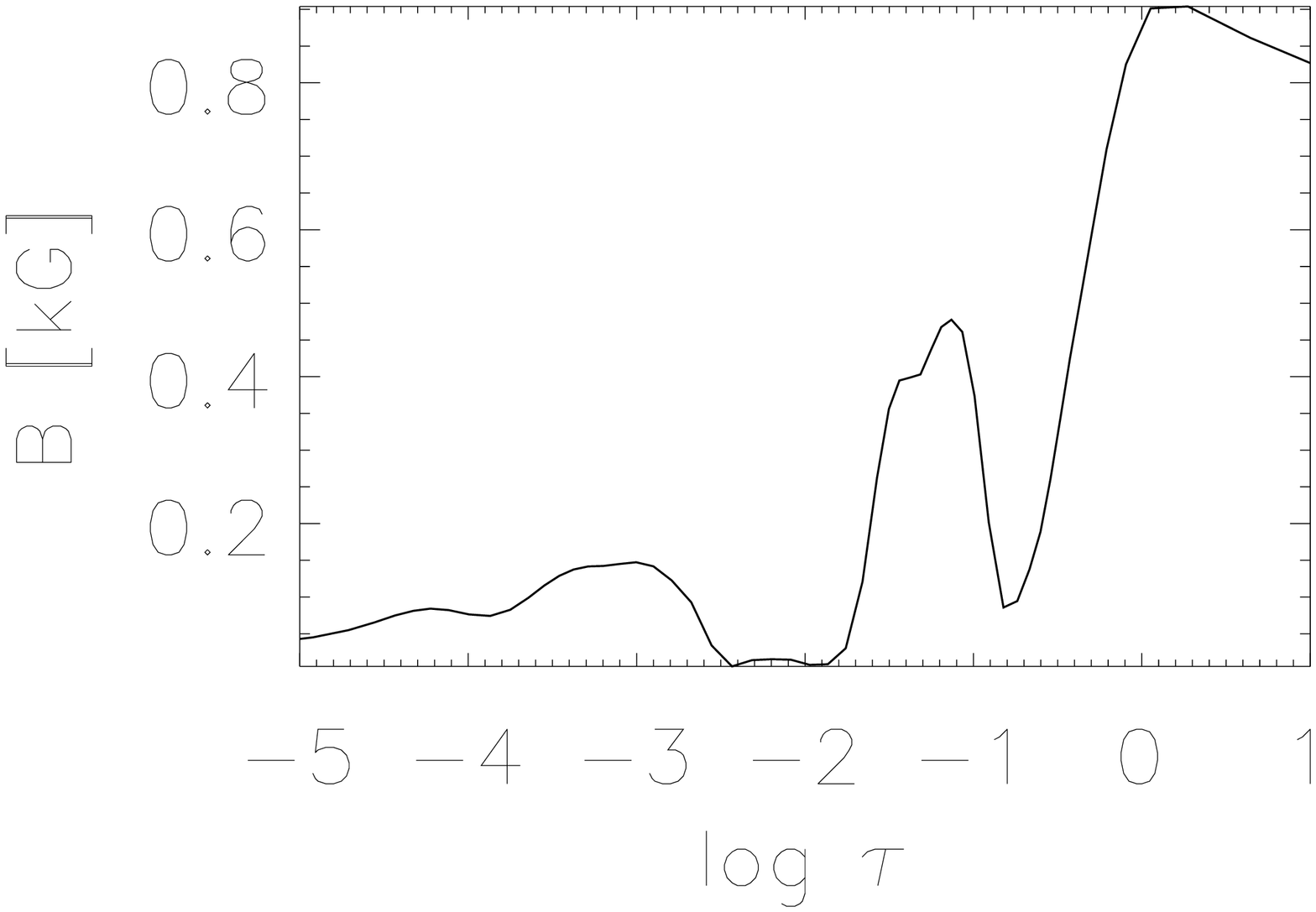}
\includegraphics*[width=0.175\linewidth,angle=90,trim= 1cm 1cm 1cm 3.3cm,clip=true]{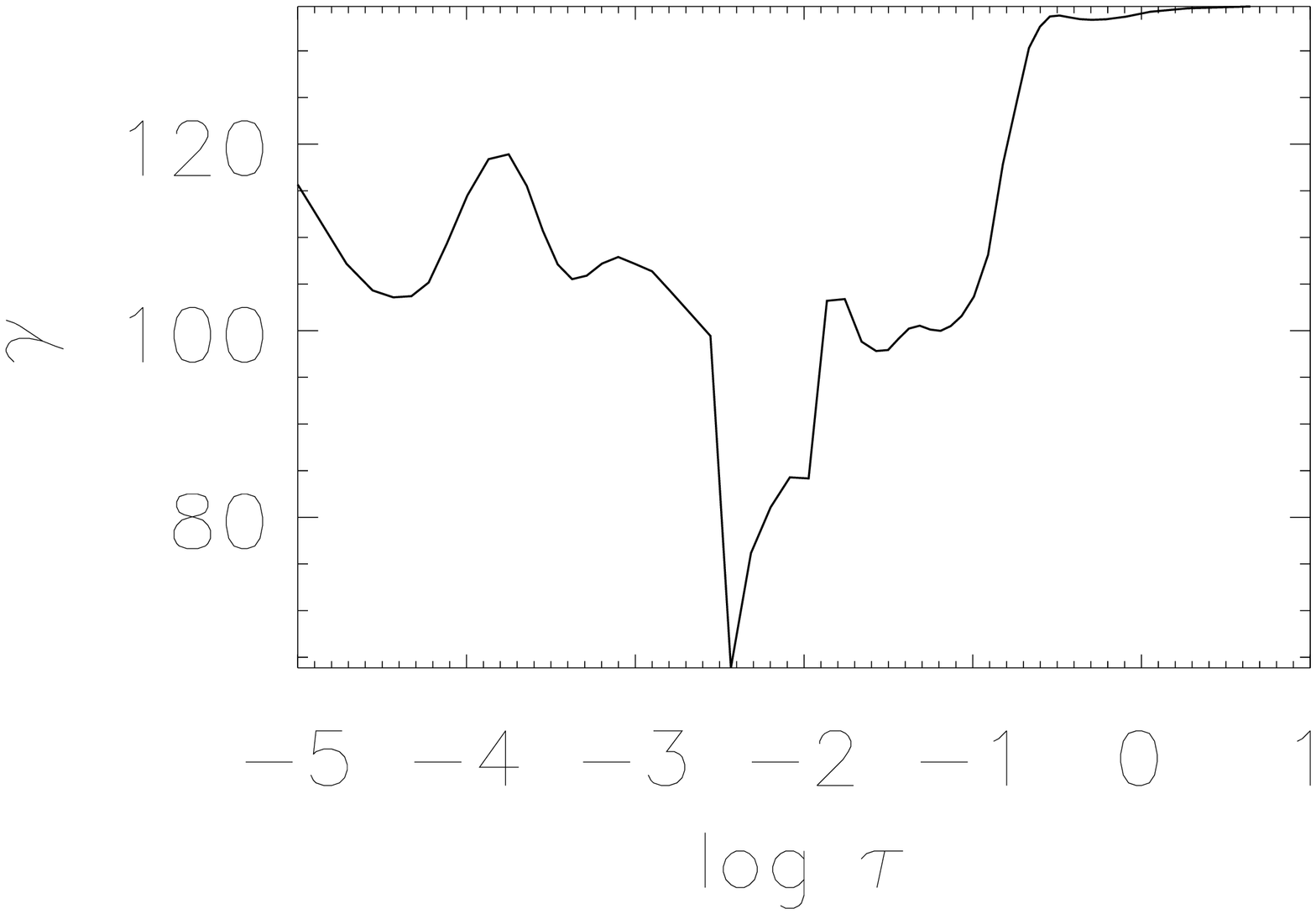}
\includegraphics*[width=0.175\linewidth,angle=90,trim= 1cm 1cm 1cm 3.5cm,clip=true]{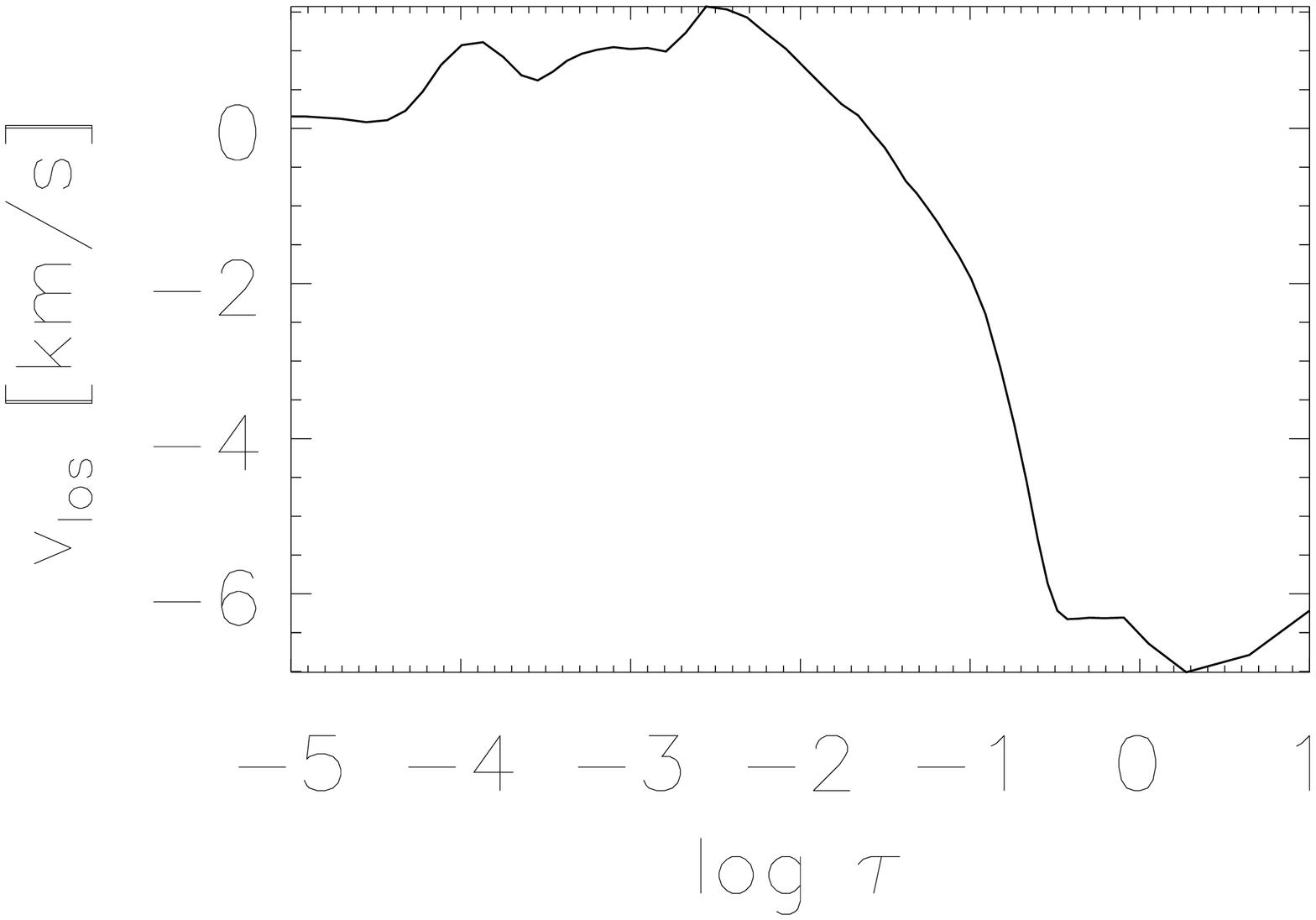}

\caption{An example of a feature that appears on the edge of a granule in the 'emergence run' at t$=156$~s. The feature is visible at $[2,2.8]$~Mm in Fig.~\ref{fig:all_imax}. \textit{Upper row, left to right} - line-of-sight (LOS) velocity, and vertical magnetic field at $\log\tau=-1$ and corresponding map of Stokes V at $\Delta\lambda=+227$~m\AA~ at original spatial resolution. The star symbols mark the position of the pixel of interest. \textit{Lower row, left to right} - Height profiles of temperature, magnetic field strength, inclinations and LOS velocity at the pixel of interest \label{fig:ex_atm}}
\end{figure*}

\begin{figure}
\centering
\includegraphics*[width=0.7\linewidth,angle=90,trim= 0cm 0cm 0cm 0cm,clip=true]{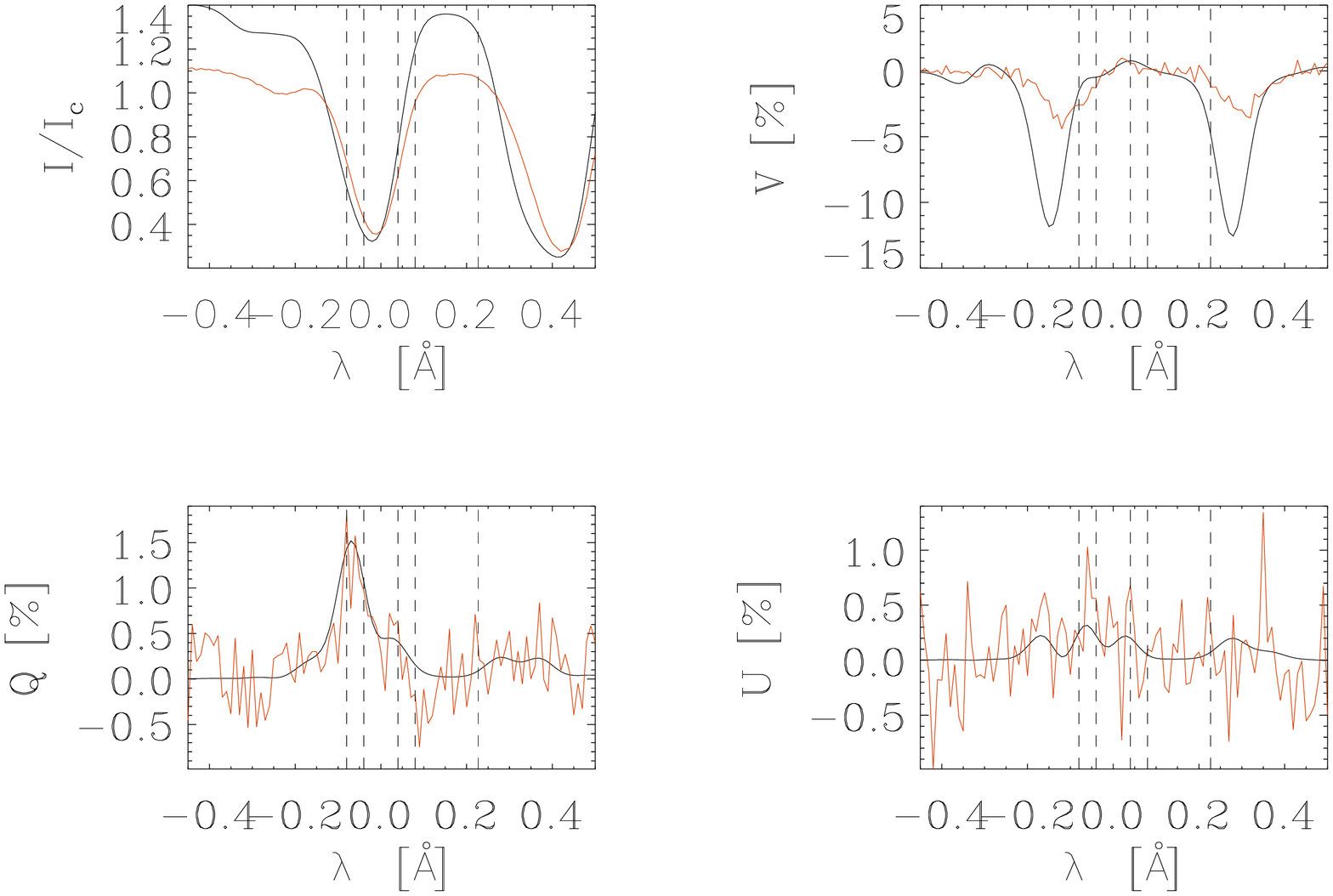} 
\includegraphics*[width=0.7\linewidth,angle=90,trim= 0cm 0cm 0cm 0cm,clip=true]{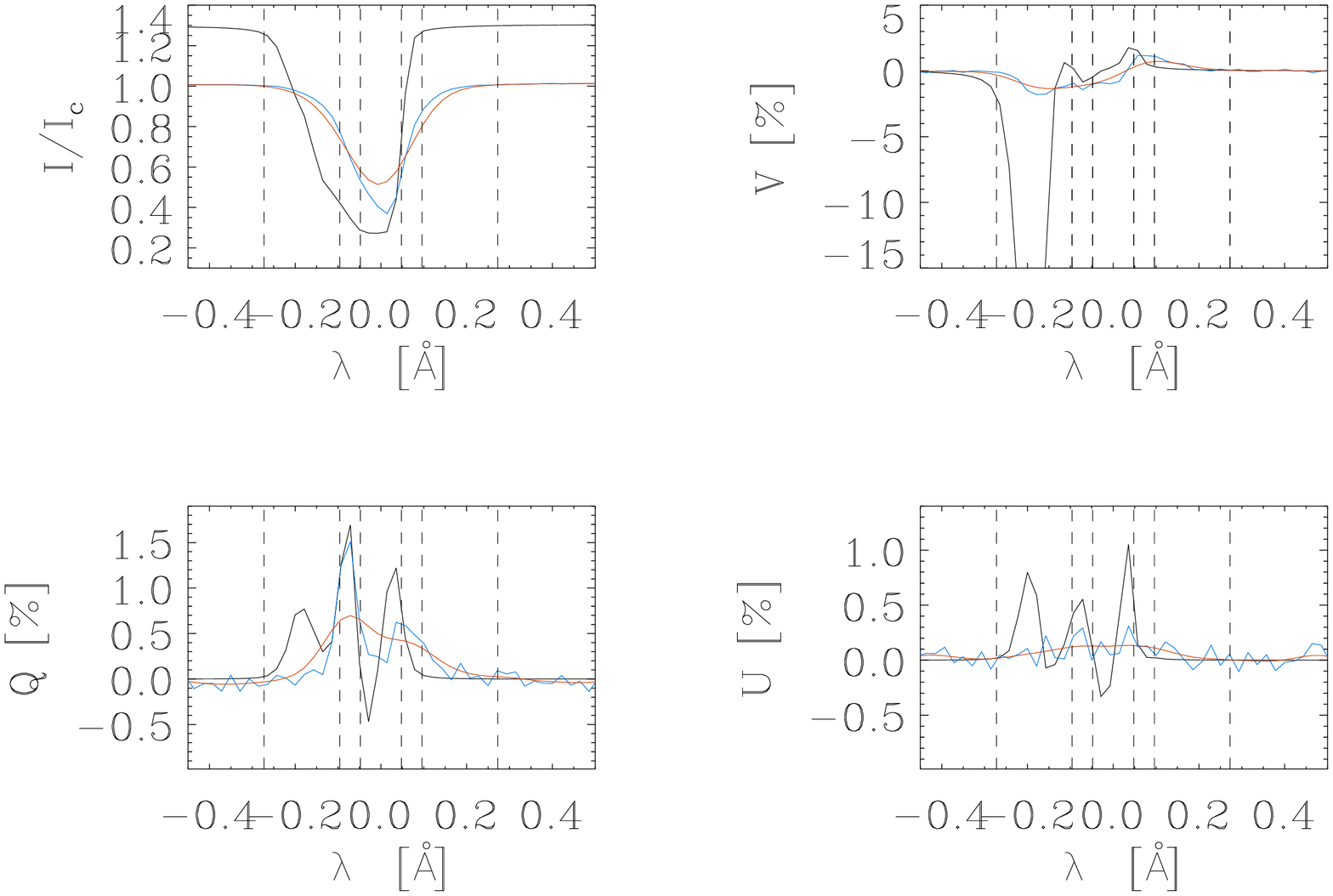} 
\caption{An example of a feature that appears on a granular edge in the 'emergence run' at t$=156$~s. Resulting Stokes I,Q,U and V profiles at the pixel shown in Fig.~\ref{fig:ex_atm}. \textit{Upper panels} - synthesized IMaX observations at the original resolution (black line) and when diffraction-limited resolution, stray light and noise are taken into account (red line); \textit{Lower panels} - synthesized Hinode observations, the Fe I 630.25~nm at original spatial resolution (black), at Hinode spatial resolution (blue) and with IMaX spectral resolution (red). \label{fig:ex_prof}}
\end{figure}

\section{Results}

The simulation without the imposed horizontal flux sheet is similar to that studied by \cite{Danilovic:etal:2010}. The main dynamical activity is the cancellation of flux between the opposite polarity magnetic elements. High velocities only occur in a few locations where several 
granules meet and magnetic field intensification takes place. Since the addition of the horizontal flux sheet leads to a substantial increase in the number of high velocity events, we will concentrate mainly on the 'emergence run'. 
Figure~\ref{fig:all_mhd} shows various plasma properties on the $< \tau_{500} > =1$ plane from this simulation at t$=156$~s.\footnote{Movies covering the $20$~minute period of both runs are available as on-line material. They combine all the maps shown in Figures~\ref{fig:all_mhd},~\ref{fig:all_imax} and ~\ref{fig:all_hinode}.} 

The flux sheet is significantly corrugated by the convective flows already $30$~s after the start. At around $90$~s, the first horizontal field crosses the $< \tau_{500} > =1$ plane. Then at $156$~s the whole simulation domain at that level is covered with horizontal magnetic field structures. Although the field strength was initially uniform in the horizontal direction, the loops are highly structured now. Fine structures in magnetic field are reflected in the fine structures in the intensity. The vertical field map shows footpoints of $\Omega$ loops as they sweep from the centres of granules towards their edges, at the same time increasing in strength as the material drains. 

As the field emerges strong upflows appear at the +X edge of the pre-existing magnetic 
elements with negative polarity and on the -X edge of positive magnetic elements. These are locations at which the emerging field has the opposite polarity to the pre-existing field. An examination of the 
magnetic field lines shows that the high velocities correspond to U loops which are formed due to reconnection 
below the surface and are accelerated outwards by magnetic tension as they cross the photosphere. This outward acceleration leads to upflows and corresponding blue-shifts of spectral lines. 
In some cases the reconnection occurs at optical depth unity where strong magnetic concentrations are pushed together by the granular flow occurs. Reconnection jets with both upflows and downflows are then present.\footnote{An example of such processes can clearly be seen in the accompanying movie around $t=406$~s at $[2.5,4.5]$~Mm.} Similar jets have previously been studied in more idealized 
simulations at different spatial scales \citep{Takeuchi:2001, Isobe08, Kigure10}.

On the other side of the flux concentrations (the -X edge of features with negative polarity and +X side of positive polarity features), the emerging flux leads not to a cancellation but to an enhancement of the field. This is consistent with the fact that the sheet of horizontal flux has zero net vertical flux.
This enhancement of the magnetic field can initiate a 'collapse' of the field (old and new) to kilo-gauss strengths. This process is aided in some cases by the draining of material down the field lines and produces strong negative vertical velocities, and consequently strong red-shifted features.\footnote{An examples of such processes can clearly be seen in the accompanying movie around $t=375$~s at $[2.3,2.8]$~Mm.}

\subsection{SUNRISE/IMaX observables}
We now consider the resulting spectropolarimetric signals as they would be seen by IMaX (Fig.~\ref{fig:all_imax}). A parameter V$_{c}$ is defined here as Stokes V signal at $\Delta\lambda=+227$~m\AA~ from Fe~I~$525.02$~nm nominal line center.
Without the imposed flux sheet, there are only few events where significantly strong V$_{c}$ appear. 
The number of such events increases significantly in the case of the 'emergence run', with the highest frequency of the events appearing in the period from t$=62$~s to t$=375$~s during which most of the horizontal field appears at the surface. The strong V$_{c}$ signals appear mostly in intergranular lanes. The locations of the strongly shifted signals change slightly as the granular motions buffet the magnetic features. The drift velocities are comparable with observations, as reported by \cite{Borrero:etal:2010}. 

The simulated V$_{c}$ features sometimes appear in pairs, e.g. the example at t$=313$~s at $[2.4,3]$~Mm in the accompanying movie. The same is seen in observations. In the simulations, these pairs of high V$_{c}$ signals coincide with locations where the mass drains and strong downflows are generated. At the same time, only very weak V$_{c}$ signals are present where upflows arise, especially after spatial smearing. Figure~\ref{fig:2dhist} displays this asymmetry between up- and downflow. We note that in general only V$_{c}>5$\% at the original spatial resolution of the simulations stays over the detection threshold after the spatial smearing. Such strong V$_{c}$ can be created only in locations where the vertical component of the magnetic field exceeds $1$~kG combined with velocities higher than $5$~km/s. The upflows that appear in the intergranular lanes coincide with strongly bent magnetic field lines, so that in  these cases the vertical component of magnetic field is rather weak. That explains why the scatter plot in Fig.~\ref{fig:2dhist} (middle panel) shows that almost all V$_{c}>5$\% are associated with downflows. The bottom panel of Fig.~\ref{fig:2dhist} reveals that only $<1$\% of pixels contains supersonic velocities. A very small percentage of those shows also V$_{c}>5$\% and those are all in downflows.

To study the statistics of these events that show high V$_{c}$ signals, we use the threshold of $|$V$_{c}|>1.25 \times 10^{-2}$~I$_{c}$ on the smeared V$_{c}$ maps, corresponding to $\approx 4 \sigma$, i.e. the same threshold as used by \cite{Borrero:etal:2010}.  Instead of using the minimum of 9 IMaX pixels, we take 3 as a minimum size of the events. In this way, we detect 3 cases in the 'reference run' and 23 cases in the 'emergence run'. Figure~\ref{fig:hist_obs} shows characteristics of these events. Comparison with Fig.~2 from \cite{Borrero:etal:2010}, which shows the corresponding histograms from the observations, reveals several interesting facts. Firstly, the simulated events are smaller and shorter lived than the observed ones (Fig.~\ref{fig:hist_obs}a and b). This is to be expected given the lower size threshold used here. Although their number is small to say it with certainty, there is a hint that the correlation between size and lifetime is present in simulations too (Fig.~\ref{fig:hist_obs}c). Thirdly, the continuum intensities in these regions are slightly lower than in observation (Fig.~\ref{fig:hist_obs}d). This is understandable because most of the events are in the intergranular lanes. The line-averaged circular polarization signal \citep[as calculated by][]{Borrero:etal:2010} is lower than the corresponding V$_{c}$ (Fig.~\ref{fig:hist_obs}e). However, the Stokes V profiles are mostly normal with an extended red lobe or additional third redshifted lobe.  Instead, if we overplot also the maximum amplitude of simulated Stokes V profiles, we find the behaviour expected for, which is that the highest signals are mostly not at $\Delta\lambda=+227$~m\AA~away from the IMaX line. It seems therefore that averaging over 4 wavelength points $\Delta\lambda=[-80, -40, 40, 80]$~m\AA~ is not a good diagnostics, since it gives a false image of the V profiles in question. Finally, the most conspicuous deference with respect to observations is visible in the LOS velocity histogram (Fig.~\ref{fig:hist_obs}e). While the majority of observed events is associated with upflows, as deduced from the Stokes I profiles, vast majority of the simulated events is associated with downflows. In the few exceptions, upflows of no more than $500$~m/s are detected.

Two of the simulated cases associated with upflows are shown in Fig.~\ref{fig:all_imax}. The atmospheric parameters for the case at location $[2,2.8]$~Mm are given in Fig.~\ref{fig:ex_atm} and the resulting line profiles in Fig.~\ref{fig:ex_prof}. The other example at $[4,3.4]$~Mm is almost the same. In both cases, the new flux emerges and where the material is more magnetized, hence more buoyant, the combination of strong upflows and the magnetopause produces single lobed Stokes V profiles with only blue lobes \citep{Steiner:2000,Dalda:etal:2012}. The maps in the upper row in Fig.~\ref{fig:ex_atm} show that strong  V$_{c}$ signal coincides with a strip of material that carries $1$~kG field and has an upflow velocity of $>7$~km/s. The height profiles (lower row in Fig.~\ref{fig:ex_atm}) show that both, magnetic field and velocity drop rapidly with height, which means that these kinds of events will appear at the moment when the newly emerging flux reaches the solar surface. The resulting profile stays single lobed with the signal above the detection limit (Fig.~\ref{fig:ex_prof}). In this case, the linear polarization signal also stays above the noise level.

\subsection{Hinode/SP observables}

In order to see if these events are visible with Hinode/SP, we follow \cite{Valentin:etal:2011}. Synthesized Hinode/SP observables are additionally spectrally smeared with Gaussian with FWHM of $95$~m\AA~and the 3 maps shown in Fig.~\ref{fig:all_hinode} are produced. The mean linear polarization maps are obtained by averaging the linear polarization signal over 4 wavelengths points [-96, -48, 48, 96]~m\AA~away from the Fe I 630.25 nm line center. The maps are somewhat different than IMaX counterparts because the IMaX line is formed higher up and tends to sample higher loops (Danilovic et al., in preparation) which combined with the higher spatial resolution of IMaX gives them stripe-like shape. Also, different noise levels are important factors as well, since most of the weak signals in the case of IMaX are lost. To identify up- and downflow events, as in \cite{Valentin:etal:2011}, the circular polarization maps V$^{-}_{c}$ and V$^{+}_{c}$ are taken at $\mp 272$~m\AA , respectively. Because of the different widths of Fe~I~$525.02$~nm and Fe~I~$630.25$~nm lines and their different sensitivity to velocities (as well as different noise levels), the strong magnetic features are prominent in both, V$^{-}_{c}$ and V$^{+}_{c}$ maps. \cite{Valentin:etal:2011} find the same in observations and call it 'leakage from magnetic network'. 

The 'emergence run' movie shows that most of the events seen in simulated IMaX observations are also visible in V$^{+}_{c}$ maps, as expected. These profiles are very asymmetric with extended red wing, or additional red lobe. On the other hand, only few events are visible in V$^{-}_{c}$. Two events appearing above the granule at $t=156$~s are right at the limit of detection if a threshold of $0.32\%$ is employed as chosen by \cite{Valentin:etal:2011}. Lower panels in Fig.~\ref{fig:ex_prof} show corresponding profiles. At the original spatial resolution, Stokes V profile is singled lobed with an amplitude of $>15\%$. After the spatial smearing it turns into a 3-lobed profiles with a blueshifted third lobe. The asymmetry in Stokes I is completely gone after the spectral smearing. And, the same as in IMaX case, Stokes Q stays above the noise.

\section{Conclusions}

Recent SUNRISE/IMaX observations revealed many short-lived events during which highly shifted circular polarization signals are detected.  Their apparent association with the appearance of linear polarization signal and the nearby opposite magnetic polarity suggest that these events might be produced by the reconnection of the emerging with the preexisting field. Here, we use realistic 3D MHD simulations to investigate the applicability of this scenario. We introduce a uniform horizontal field $\approx 300$~km below the solar surface into a mixed polarity run and search for counterparts to the observed signatures. We simulate SUNRISE/IMaX and Hinode/SP observables. The simulations show that events similar to the observations tend to appear much more often when emergence is taking place and only in the locations where the high line-of-sight velocities coincide with the strong longitudinal magnetic field. These simulated events qualitatively match the observations. The synthesized line profiles are very similar to the observed ones. However there are a few significant deferences. The simulated events are on average considerably smaller and weaker than observed. For the mean lifetime and size of these events we obtain $57.4$~s and $7.6$~pixels, respectively, instead of $81.3$~s and $15.5$~pixels as found by \cite{Borrero:etal:2010} in the observations. Also, the huge majority of them appears in intergranular lanes and are associated with downflows, in contrast to observations. We believe that the size limit introduced by \cite{Borrero:etal:2010} removed most of these events. As a result, their analyses included only the cases that appear above granule. We find a few of these in our simulations. They are produced at the moment when the newly emerging flux appears at the surface. At these locations both, magnetic field and velocity drop rapidly with height. We find no change of polarity along the line of sight, as suggested by \cite{Borrero:etal:2013}. Besides, our simulation give none of the height profiles presented in that paper. In the simulations, reconnection happens in the intergranular lanes where magnetic features of opposite polarity come into contact. Depending on the gradient of magnetic field, the temperature height profile can show a bump corresponding to an increase of  $\approx 1000$~K. As a result, a shift in the formation height of the synthesized Fe I lines is produced. Because of this, lines usually miss to reveal dramatic events or show only associated downflows which take place below the temperature bump i.e. the reconnection site.

Why couldn't we produce more of the observed-like examples? Having in mind that these features appear rather seldom, only few over 55$^{\prime\prime}\times$55$^{\prime\prime}$ \citep{Valentin:etal:2011}, this might not be so surprising. Thus, simulations that cover larger field of view  and extend deeper in the convention zone are needed. We expect that such simulations will result in more realistic emergence events and wider spectra of magnetic field configurations that are not imposed by initial and boundary conditions as in our case.


\section{Appendix}

\begin{figure*}
\centering
\includegraphics[width=0.6\linewidth,angle=-90,trim= 0cm 0cm 0cm 0cm,clip=true]{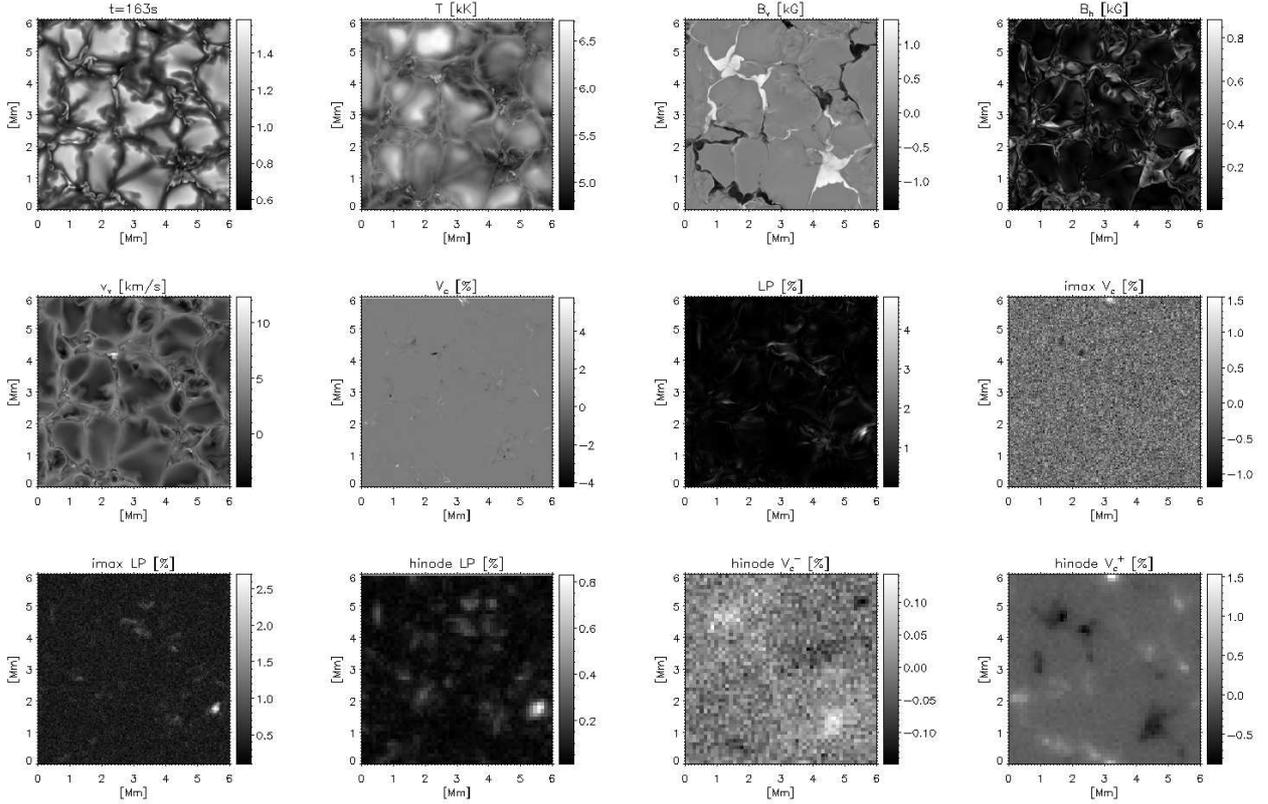}
\caption{A snapshot from the 'reference' run. \textit{From left to the right and from top to the bottom}: bolometric intensity; temperature, vertical and horizontal field, line-of-sight velocity (upflow is negative) at constant geometrical height that corresponds to $\langle\tau_{500}\rangle\approx~1$; Stokes V $\Delta\lambda=+227$~m\AA~away from nominal Fe I $525.02$~nm line center and mean linear polarization at original and IMaX resolution; mean linear polarization and Stokes V at $\Delta\lambda=-227$~m\AA~and $+227$~m\AA~away from nominal Fe I $630.25$~nm line center.  \label{fig:reference}}
\end{figure*}

\begin{figure*}
\centering
\includegraphics[width=0.6\linewidth,angle=-90,trim= 0cm 0cm 0cm 0cm,clip=true]{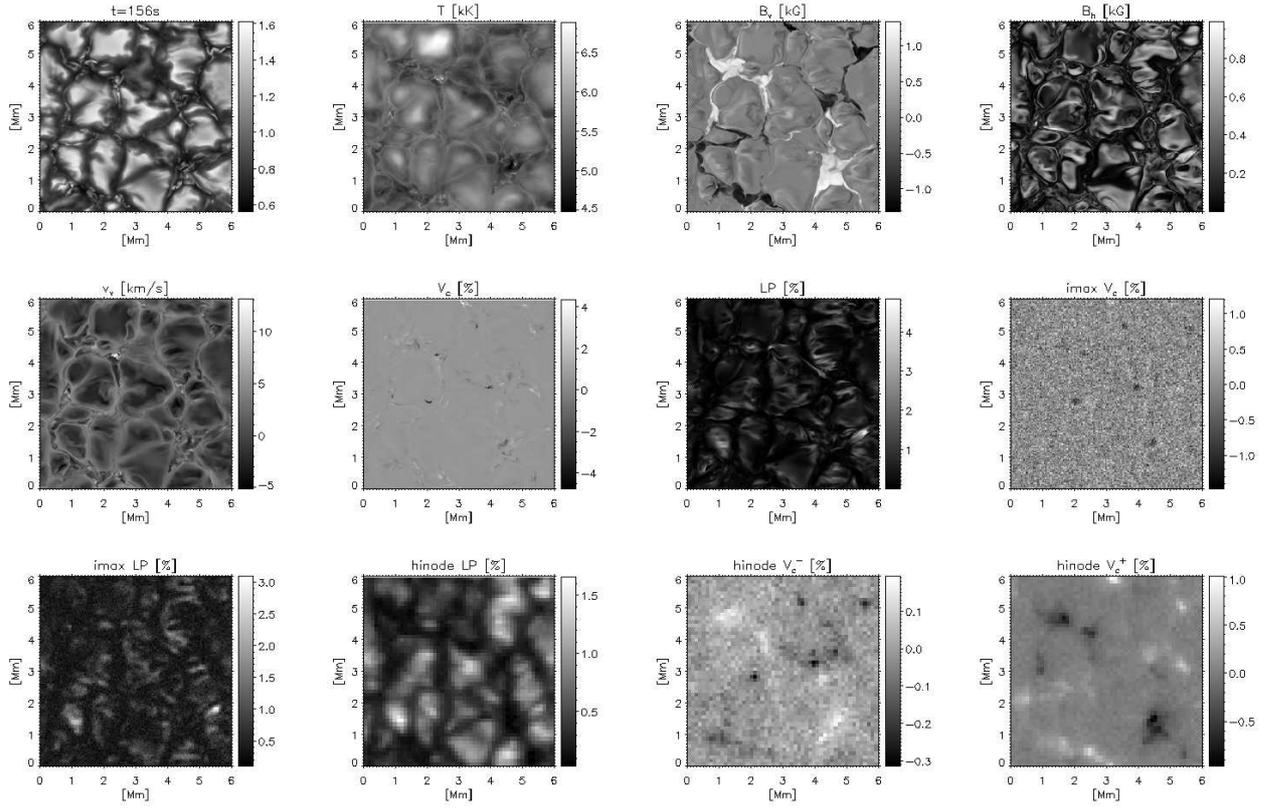}
\caption{A snapshot from the 'emergence' run. The same as in Fig.~\ref{fig:reference}.   \label{fig:emergence}}
\end{figure*}

Two movies for both the 'reference' and the 'emergence' runs are attached to the paper. Figures ~\ref{fig:reference} and ~\ref{fig:emergence} show the styles from the movies.

\end{document}